\documentclass{article}

\usepackage{amssymb}

\begin{document}

\title{\bf Localization of gravity in brane world with arbitrary extra
dimensions }
\author{ A. M. Yazdani$^1$, K. Atazadeh$^{2}$ \hspace{.1 cm} and \hspace{.1 cm} S.
Jalalzadeh$^{2}$\thanks{email:
s-jalalzadeh@sbu.ac.ir}
\\ $^2${\small Department of Physics, Shahid Beheshti University G. C., Evin, Tehran 19839, Iran}
\\ $^1${\small Department of Physics, Qom Branch, Islamic Azad University, Qom, Iran}
}

\maketitle

\begin{abstract}
We study the induced $4$-dimensional linearized Einstein field
equations in an $m$-dimensional bulk space by means of a confining
potential. We used the confining potential in this model to localized gravitons
on the brane. It is shown that in this approach the mass of graviton is
quantized. The cosmological constant problem is also addressed
within the context of this approach. We show that the difference
between the values of the cosmological constant in particle physics
and cosmology stems from our measurements in two different scales,
small and large.
\vspace{5mm}\\
pacs: {04.50.+h, 04.30.-w}
\end{abstract}

\maketitle

\section{Introduction}
The view that our universe might actually have dimensions more than
four is anchored on the recent developments in string and M-theories
in which gravity arises as a truly higher dimensional. Only in the
low energy limit it manifests in the familiar 4-dimensional general
relativity. Recent observations have indicated that the cosmological
constant $\Lambda$ is a sufficiently small positive value,
$\Lambda\sim(10^{-3}eV)^{4}$ \cite{1}. Consequently it is natural to
assume that we live in de a Sitter universe. However it is difficult
to explain the reason for the smallness of $\Lambda$, the so called
cosmological constant problem. By using the warped extra dimension
scenario in \cite{2}, new suggestion for the cosmological constant
problem was proposed. Randall and Sundrum investigated that the
gravity can be localized on flat 3-brane embedded in $AdS_{5}$, the
usual 4-dimensional gravity can be reproduced at large distance
scale. This is because a normalizable zero mode exists and the
effective four-dimensional Planck scale is finite. Due to
fine-tuning between the bulk cosmological constant and brane
tension, the four-dimensional cosmological constant is vanishing.
Furthermore, the close relation between $AdS$/CFT correspondence and
the Randall-Sundrum model is explored in the last couple of years
\cite{3,4,5}. Moreover it is expected that the warped braneworlds
provide a new scenario in the framework of phenomenological model
building and cosmology. Based on the assumption that physics in four
dimensions should be explained by higher dimensional theory, the
warped braneworlds with an infinite extra dimension are remarkable
scenarios in the field of particle physics.

There are three usual models to consider when studying the dynamics of brane-worlds models featuring boundary terms in the action
and sometimes mirror symmetry, such that bulk gravitational waves interfere with the brane-world
motion. 
The first model usually comes together with junction conditions producing an algebraic relationship
between the extrinsic curvature and the confined matter \cite{isreal}.
In this  model the problem of localization of gravity is discussed in
several papers. For example, in \cite{11}, the authors addressed the
localization of gravity in a FRW type brane
embedded in either $AdS_5$ or $dS_5$ bulk space, and  show that
the graviton zero mode is trapped on the brane. Non-trapped, massive
Kaluza-Klein (KK) modes correspond to a correction to Newton's law.
Also in \cite{12} they consider Randall-Sundrum model with localized
gravity, replacing the extra compact space-like dimension by a
time-like one. In this way they show that the solution to the
hierarchy problem can be reconciled with the correct cosmological
expansion of the visible universe. One the other hand, In \cite{13},
the authors show that, in brane models with a compact extra dimension
like the Randall-Sundrum's two brane model or models with universal extra
dimension \cite{14}, which allow a zero mode  known from
KK theories, the five dimensional linearly perturbed Einstein
equations are in conflict with observations. 

The second model is against the Israel junction conditions for confining matter or gauge fields on the brane by using $Z_2$ symmetry.
This model was introduced in \cite{6} where matter
fields are trapped on a 4-dimensional hypersurface by the action of
the confining potential ${\cal V}$.
In this model, without using $Z_2$ symmetry or without postulating any junction condition,
Friedman equation is modified by a geometrical term which is defined in terms of the extrinsic
curvature, leading to a geometrical interpretation for dark energy and to find a richer set
of cosmological solutions in accordance with the current observations \cite{maia,maia2}.
In this direction, in \cite{7} the authors developed a brane model, in which the number of
non-compact extra dimensions is arbitrary also the dynamics of test particles confined
to a brane by the action of such potential at the classical and quantum
levels were studied in \cite{7}. In \cite{8}, a brane-world model was studied in which matter is confined to
the brane through the action of such a potential without using any junction conditions, offering a
geometrical explanation for the accelerated expansion of the universe.

Third model for localizing gravity on the brane was introduced in \cite{maia4} where the foundation of model based on a geometrical hypothesis. 
In \cite{maia4} the authors investigated the most general geometrical scenarios
in which a brane-world program compatible with
the hypotheses of embedding, confinement and the existence
of quantum states can be implemented and showed that their analysis
is independent of any previous choice of geometry,
topology, number of dimensions and signature for the
bulk 
There have also been arguments concerning the uniqueness of the junction conditions. Indeed, other forms of junction conditions
exist, so that different conditions may lead to different physical results \cite{battye}. Furthermore, these
conditions cannot be used when more than one non-compact extra dimension is involved.

In this paper, we follow \cite{7,8} and consider an $m$-dimensional bulk space without
imposing any junction condition.  Instead we used confining potential approach that
localized gravitons according to the equation (\ref{1-40}) on the brane. Finally we argue that the value of mass of graviton
that in this model is quantized.

\section{The model}
In this section we present a brief review of the model proposed in
\cite{7,8}.  Consider the background manifold $\overline{V}_{4}$
isometrically embedded in a pseudo-Riemannian manifold $V_{m}$ by
the map ${ \cal Y} : \overline{V}_{4}\rightarrow  V_{m}$ such that
\begin{eqnarray}\label{1-1}
{\cal G} _{AB} {\cal Y}^{A}_{,\mu } {\cal Y}^{B}_{,\nu}=
\bar{g}_{\mu \nu}  , \hspace{.5 cm} {\cal G}_{AB}{\cal
Y}^{A}_{,\mu}{\cal N}^{B}_{a} = 0,\hspace{.5 cm}  {\cal G}_{AB}{\cal
N}^{A}_{a}{\cal N}^{B}_{b} = g_{ab}= \pm 1.
\end{eqnarray}
where $ {\cal G}_{AB} $  $ ( \bar{g}_{\mu\nu} ) $ is the metric of
the bulk (brane) space  $  V_{m}  (\overline{V}_{4}) $ in arbitrary
coordinates, $ \{ {\cal Y}^{A} \} $   $  (\{ x^{\mu} \}) $  is the
basis of the bulk (brane) and  ${\cal N}^{A}_{a}$ are $(m-4)$ normal
unite vectors, orthogonal to the brane. The deformation of
$\bar{V}_{4}$ in a sufficiently small neighborhood of the brane
along an arbitrary transverse direction $\zeta$ is given by
\begin{eqnarray}\label{1-2}
{\cal Z}^{A}(x^{\mu},\xi^{a}) = {\cal Y}^{A} + ({\cal
L}_{\zeta}{\cal Y})^{A}, \label{eq2}
\end{eqnarray}
where $\cal L$ represents the Lie derivative. By choosing
$\zeta^{a}$ orthogonal to the brane, we ensure gauge independency
\cite{7} and  have deformations of the embedding along a single
orthogonal extra direction $\bar{{\cal N}}_{a}$ giving local
coordinates of the deformed brane as
\begin{eqnarray}\label{1-3}
{\cal Z}^{A}_{,\mu}(x^{\nu},\xi^{a}) = {\cal Y}^{A}_{,\mu} +
\xi^{a}\bar{{\cal N}}^{A}_{a,\mu}(x^{\nu}),
\end{eqnarray}
where $\xi^{a}$ $(a = 1,2,...,m-4)$ is a small parameter along
${\cal N}^{A}_{a}$ that parameterizes the extra noncompact
dimensions. One can see from equation (\ref{1-2}) that since the
vectors $\bar{{\cal N}}^{A}$ depend only on the local coordinates
$x^{\mu}$, they do not propagate along the extra dimensions
\begin{eqnarray}\label{1-4}
{\cal N}^{A}_{a}(x^{\mu}) = \bar{{\cal N}}^{A}_{a} +
\xi^{b}[\bar{{\cal N}}_{b} , \bar{{\cal N}}_{a}]^{A} = \bar{{\cal
N}}^{A}_{a}.
\end{eqnarray}
The above  assumptions lead to the embedding equations of the
deformed geometry
\begin{eqnarray}\label{1-5}
{\cal G}_{\mu \nu }={\cal G}_{AB}{\cal Z}_{\,\,\ ,\mu }^{A}{\cal
Z}_{\,\,\ ,\nu }^{B},\hspace{0.5cm}{\cal G}_{\mu a}={\cal
G}_{AB}{\cal Z}_{\,\,\ ,\mu }^{A}{\cal N}_{\,\,\ a}^{B},\hspace{.5 cm}{\cal G}_{ab}= {\cal G}_{AB}{\cal N}_{\,\,\ a}^{A}%
{\cal N}_{\,\,\ b}^{B}.
\end{eqnarray}
If we set ${\cal N}_{\,\,\ a}^{A}=\delta _{a}^{A}$, the metric of
the bulk space can be written in the following matrix form
\begin{equation}
{\cal G}_{AB}=\left( \!\!\!%
\begin{array}{cc}
g_{\mu \nu }+A_{\mu c}A_{\,\,\nu }^{c} & A_{\mu a} \\
A_{\nu b} & g_{ab}%
\end{array}%
\!\!\!\right) ,  \label{1-6}
\end{equation}%
where
\begin{equation}
g_{\mu \nu }=\bar{g}_{\mu \nu }-2\xi ^{a}\bar{K}_{\mu \nu a}+\xi ^{a}\xi ^{b}%
\bar{g}^{\alpha \beta }\bar{K}_{\mu \alpha a}\bar{K}_{\nu \beta b},
\label{1-7}
\end{equation}%
is the metric of the deformed brane, so that
\begin{equation}
\bar{K}_{\mu \nu a}=-{\cal G}_{AB}{\cal Y}_{\,\,\,,\mu }^{A}{\cal
N}_{\,\,\ a;\nu }^{B},  \label{1-8}
\end{equation}%
represents the extrinsic curvature of the original brane (second
fundamental form). Also, we use the notation $A_{\mu c}=\xi
^{d}A_{\mu cd}$, where
\begin{equation}
A_{\mu cd}= {\cal G}_{AB}{\cal N}_{\,\,\ d;\mu }^{A}{\cal N}_{\,\,\ c}^{B}=%
\bar{A}_{\mu cd},  \label{1-9}
\end{equation}%
represents the twisting vector fields (the normal fundamental form). Any fixed $%
\xi ^{a}$ signifies a new deformed geometry, enabling us to define
an extrinsic curvature similar to the original one by
\begin{eqnarray}
\widetilde{K}_{\mu \nu a}=-{\cal G}_{AB}{\cal Z}_{\,\,\ ,\mu }^{A}{\cal N}%
_{\,\,\ a;\nu }^{B}= \bar{K}_{\mu \nu a}- \xi ^{b}\left( \bar{K}_{\mu \gamma a}%
\bar{K}_{\,\,\ \nu b}^{\gamma }+A_{\mu ca}A_{\,\,\ b\nu }^{c}\right)
. \label{1-10}
\end{eqnarray}%
Note that definitions (\ref{1-7}) and (\ref{1-10}) require
\begin{equation}
\widetilde{K}_{\mu \nu a}=-\frac{1}{2}\frac{\partial {\cal G}_{\mu \nu }}{%
\partial \xi ^{a}}.  \label{1-11}
\end{equation}%
In geometric language, the presence of gauge fields $A_{\mu a}$
tilts the embedded family of sub-manifolds with respect to the
normal vector ${\cal N} ^{A}$. According to our construction, the
original brane is orthogonal to the normal vector ${\cal N}^{A}.$
However,  equation (\ref{1-5})  shows that this is not true for the
deformed geometry. Let us change the embedding coordinates and set
\begin{equation}\label{1-12}
{\cal X}_{,\mu }^{A}={\cal Z}_{,\mu }^{A}-g^{ab}{\cal
N}_{a}^{A}A_{b\mu }.
\end{equation}%
The coordinates ${\cal X}^{A}$ describe a new family of embedded
manifolds whose members are always orthogonal to ${\cal N}^{A}$. In
this coordinates the embedding equations of the deformed brane is
similar to the original one, described by equations (\ref{1-1}), so
that ${\cal Y}^{A}$ is replaced by ${\cal X}^{A}$. This new
embedding of the local coordinates are suitable for obtaining
induced Einstein field equations on the brane. The extrinsic
curvature of a deformed brane then becomes
\begin{eqnarray}
K_{\mu \nu a}=-{\cal G}_{AB}{\cal X}_{,\mu }^{A}{\cal N}_{a;\nu }^{B}=\bar{K}%
_{\mu \nu a}-\xi ^{b}\bar{K}_{\mu \gamma a}\bar{K}_{\,\,\nu
b}^{\gamma } = - \frac{1}{2}\frac{\partial g_{\mu \nu
}}{\partial \xi ^{a}}, \label{1-13}
\end{eqnarray}%
which is the generalized York relation and shows how the extrinsic
curvature propagates as a result of the propagation of the metric in
the direction of extra dimensions.
Now, let us write the Einstein equations in the bulk space as
\begin{equation}\label{1-14}
G^{(b)}_{AB}+\Lambda^{(b)} {\cal G}_{AB}=-\alpha^{*} S_{AB},
\end{equation}
where $\alpha^{*}=\frac{1}{M_{*}^{m-2}}$ ($M_{*}$ is the
fundamental scale of energy in the bulk space), $\Lambda^{(b)}$ is
the cosmological constant of the bulk and $S_{AB}$ consists of two
parts
\begin{equation}\label{1-15}
S_{AB}=T_{AB}+ \frac{1}{2} \cal{V} G_{AB},
\end{equation}
where $T_{AB}$ is the energy-momentum tensor of the matter confined
to the brane through the action of the confining potential
$\cal{V}$ \cite{7,8}. Using the Einstein equations (\ref{1-14}) and
Gauss-Codazzi equations, the tangent component of equation
(\ref{1-14}) becomes (gravi-tensor equation) \cite{8}
\begin{eqnarray}\label{1-16}
G_{\mu\nu} -(\frac{m-7}{m-1})\Lambda^{(b)}g_{\mu\nu} = Q_{ \mu\nu} - {\cal
E}_{\mu\nu} -\frac{2}{m-2}\alpha^{*}S_{\mu\nu} \nonumber \\ -
\left(\frac{m-3}{m-2}\right)\alpha^*g^{ab}S_{ab}g_{\mu\nu} +
\frac{(m-4)(m-3)}{(m-2)(m-1)}\alpha^*Sg_{\mu\nu} ,
\end{eqnarray}
where ${\cal E}_{\mu\nu} = g^{ab}{\cal C}_{ABCD}{\cal N}^A_a {\cal
X}^B_{\mu}{\cal X}^C_{\nu}{\cal N}^D_b$ is the electric part of Weyl
tensor of the bulk space ${\cal C}_{ABCD}$, and
\begin{eqnarray}
Q_{\mu\nu} = g^{ab}\left(K_{a\mu\gamma}K_{b\nu}^{\,\,\, \gamma} -
K_aK_{b\mu\nu}\right)-\frac{1}{2}\left(K_{a\alpha\beta}K^{a\alpha\beta} - K^aK_a
\right)g_{\mu\nu}.\label{1-16-1}
\end{eqnarray}
On the other hand, again from (\ref{1-14}), the trace of Codazzi's
equation gives the gravi-vector equation
\begin{eqnarray}
K^\delta_{a\gamma;\delta} - K_{a,\gamma} - A_{ba\gamma}K^b + A_{ba\delta}K^{b\delta} +  B_{a\gamma} =
\frac{3(m-4)}{m-2}\alpha^{*}S_{a\gamma},\label{1-17}
\end{eqnarray}
where
\begin{equation}
B_{a\gamma} = g^{mn}C_{ABCD}{\cal N}^A_m{\cal N}^B_a{\cal
X}^C_{,\gamma}{\cal N}^D_n.\label{1-18}
\end{equation}
Finally, gravi-scalar equation is given by \cite{8}
\begin{eqnarray}
\alpha^{*}\left[\frac{m-5}{m-1}S - g^{mn}S_{mn}\right]g_{ab} = \frac{m-2}{6}\left(Q + R +W\right)g_{ab} -
\frac{4}{m-1}\Lambda^{(b)}g_{ab},\label{1-19}
\end{eqnarray}
where
\begin{equation}
W = g^{ab}g^{mn}C_{ABCD}{\cal N}^A_m{\cal N}^B_b{\cal N}^C_n{\cal
N}^D_a.\label{1-20}
\end{equation}
Equations (\ref{1-16})-(\ref{1-20}) represents the Einstein field
equations on the bulk space near the original brane, represented in
the Gaussian coordinates. The confinement hypotheses is applied as
an effect of confinement potential on the original brane, and can be
implemented simply as \cite{8}
\begin{equation}
\alpha\tau_{\mu\nu} =
\frac{2}{(m-2)}\alpha^{*}\bar{T}_{\mu\nu}, \hspace{.5
cm}\bar{T}_{\mu a}=0, \hspace{.5 cm}\bar{T}_{ab}=0,\label{1-21}
\end{equation}
where $\alpha$ is the scale of energy on the brane. Now, the induced
Einstein field equation on the original brane can be written
\begin{eqnarray}
G_{\mu\nu}+ \Lambda g_{\mu\nu} =- \alpha \tau_{\mu\nu}
+\frac{(m-4)(m-3)}{2(m-1)}\alpha\tau g_{\mu\nu} +  Q_{\mu\nu} - {\cal
E}_{\mu\nu},\label{1-22}
\end{eqnarray}
where  $\Lambda=- \frac{m-7}{m-1} \Lambda^{(b)}$. As was mentioned before,
$Q_{\mu\nu}$ is a conserved quantity which according to \cite{maia2} may
be considered as an energy-momentum tensor of a dark energy fluid
representing the $x$-matter, the more common phrase being
``X-Cold-Dark Matter''. This matter has the most general form of the
equation of state which is characterized by the following conditions
\cite{10}: first it violates the strong energy condition at the
present epoch for $\omega_x<-1/3$ where $p_x=\omega_x\rho_x$,
second, it is locally stable {\it i.e.} $c^2_s=\delta
p_x/\delta\rho_x\ge 0$, and third, causality holds good, that is
$c_s\le 1$. Ultimately, we have three different types of ``matter''
on the right hand side of equation (\ref{1-16}), namely, ordinary
confined conserved matter represented by $\tau_{\mu\nu}$,
the XCDM matter represented by $Q_{\mu\nu}$ and finally, the Weyl matter ${\cal E}_{\mu\nu}$.\\
\section{Linearization}
To discuss linearized gravity in this approach, consider the linear
expansion of the bulk metric around the $mD$ background spacetime
\begin{equation}\label{1-23}
{\cal G}_{AB} =\hat{{\cal G}}_{AB}+ h_{AB}.
\end{equation}
where $\hat{{\cal G}}_{AB}$ is the bulk space background metric and
$h_{AB}$ is the deformation of the metric. Let us perturb the
Einstein equations of motion (\ref{1-14}) around the $\hat{{\cal
G}}_{AB}$. The resulting field equation is
\begin{equation}\label{1-24}
\delta G_{AB}^{(b)}=(\frac{1}{2}\alpha^{*}{\cal
V}-\Lambda^{(b)})h_{AB},
\end{equation}
where $\delta G_{AB}^{(b)}$ is the perturbation of the bulk space
Einstein tensor which is given by \cite{15}
\begin{equation}\label{1-25}
\delta G_{AB}^{(b)}=\tilde{G}_{AB}^{(a)}+\Psi_{AB},
\end{equation}
where
\begin{eqnarray}\label{1-26}
\tilde{G}_{AB}^{(a)}=\frac{1}{2}\left(\Box
h_{AB}-{h_{C(A;B)}}^{;C}+h_{;AB}- {\cal
G}_{AB}\left(\Box h-h^{CD}_{\,\,\,\,\,\,;CD}\right)\right),
\end{eqnarray}
and
\begin{equation}\label{1-27}
\Psi_{AB}=\frac{1}{2}\left(\hat{{\cal G}}_{AB}\hat{{\cal
R}}^{CD}h_{CD}-\hat{{\cal R}}h_{AB}\right).
\end{equation}
where $\hat{{\cal R}}^{CD}$ and $\hat{{\cal R}}$ are associated with
$\hat{{\cal G}}_{AB}$. It is straightforward to show that the linear
approximation of Einstein field equation in the bulk space, in the
harmonic gauge $h^{A}_{\,\,A}=0$ and $h^{A}_{\,\,B;A}=0$, for de
Sitter brane is
\begin{eqnarray}\label{1-28}
\Box h_{AB} - \hat{{\cal G}}_{AB}\hat{{\cal
R}}^{CD}h_{CD}-\hat{{\cal R}} h_{AB} = \alpha^{*}{\cal
V}h_{AB}-2\Lambda^{(b)}h_{AB}.
\end{eqnarray}
Inserting extra conditions  $h_{ab}=h_{a\mu}=0$ \cite{16}, we can
now write the induced linearized field equation on the  perturbed
brane
\begin{eqnarray}\label{1-29}
\Box h_{\mu\nu}+\frac{2}{m-2}\left(\alpha^*{\cal V}
-\frac{m-1}{m-7}\Lambda \right)h_{\mu\nu}=0.
\end{eqnarray}
To obtain field equations on the original brane which according to
 \cite{7} describe the physical $4D$ universe in large scales, we
expand the  d\'{}Alembertian and confining potential around the
original geometric configuration. In general we can expand the confining
potential in terms of extra dimensions $\xi^a$
\begin{equation}
{\cal V} = {\cal V}_0 + \xi^a {\cal V}_{,a} +\frac{1}{2!}\xi^a \xi^b
{\cal V}_{,ab} + {\cal O}(\xi^3),\label{1-30}
\end{equation}
where ${\cal V}_0$ is a constant that can be absorbed in
cosmological constant. To preserving the symmetries of the induced
gauge fields, we assume that ${\cal V}_{,ab} = \omega^2g_{ab}$
\cite{7}. On the other hand, to cancel the higher order expansion in
the equation (\ref{1-30}), we consider $\omega$ to be much larger
than the inverse of the scale of curvatures $\rho^{-1}$ on the
original brane, or more specifically $\omega \gg \rho^{-2}$
\cite{yaf}. This means the effective size of extra dimensions
through which the  graviton can propagate, is more smaller than the
distance of horizon denoted by $\rho$. Now, we adsorb the scale of
$\omega$ into a small dimensionless parameter $\sigma$, that is
$\omega \rightarrow \omega/\sigma$, so that $\omega$ becomes of the
same order as $\rho^{-2}$. In fact, the ``squeezing'' parameter
$\sigma$ plays the role of natural perturbation parameter. To
identify ${\cal V}_{,a}$, let us attend to the conservation equation
of $S^{AB}$ in the bulk space. From Einstein field equations
(\ref{1-14}) we have
\begin{equation}
S^{AB}_{\,\,\,\,\,\, ;A}=0.\label{1-31}
\end{equation}
Now with an eye on the confinement hypotheses (\ref{1-21}), we have
for the extra-extra components of equation (\ref{1-31}) on the
original brane
\begin{equation}
\Gamma^\mu_{\mu
c}T^{cb}+\Gamma^b_{\mu\nu}T^{\mu\nu}+\Gamma^a_{ac}T^{cb}+\frac{1}{2}{\cal
V}^{,b}=0,\label{1-32}
\end{equation}
where $\Gamma$ denotes the connection coefficients on the bulk
space. Using relation of connection coefficients on the bulk with
induced quantities obtained in \cite{jalal2}, equation (\ref{1-32})
becomes
\begin{equation}
\bar{K}^a_{\mu\nu}\bar{T}^{\mu\nu} = {\cal V}^{,a}.\label{1-33}
\end{equation}
Hence confining dictates
\begin{equation}
{\cal V}^{,a} = \bar{K}^a_{\mu\nu}\bar{T}^{\mu\nu} =
\frac{m-2}{2}\frac{\alpha}{\alpha^*}\bar{K}^a_{\mu\nu}\tau^{\mu\nu}.\label{1-34}
\end{equation}
Also, using the explicit form of the Gaussian metric (\ref{1-12})
the d\'{}Alembertian operator on the deformed brane appeared in
equation (\ref{1-29}) can be written as
\begin{eqnarray}\label{1-35}
\Box = -\frac{1}{|g|^{1/4}}\partial_{a}|g|^{1/2}\partial^{a}\frac{1}{|g|^{1/4}}
-\frac{1}{|\bar{g}|^{1/4}|g|^{1/4}}\partial_{\mu}g^{\mu\nu}|g|^{1/4}\partial_{\nu}\frac{|\bar{g}|^{1/4}}{|g|^{1/4}}.
\end{eqnarray}
Since $\sigma$ is the perturbation parameter, by changing the extra
coordinates as $\xi^a \rightarrow \sigma^{1/2}\xi^a$, the
d\'{}Alembertian  and confining potential can be expanded in powers
of $\sigma$, leading to
\begin{eqnarray}\label{1-36}
\Box =\frac{1}{\sigma} \Box^{(0)} + \Box^{(1)} +\cdots,
\end{eqnarray}
and
\begin{eqnarray}
{\cal V}
=\frac{m-2}{2}\frac{\alpha}{\alpha^*}\sigma^{\frac{1}{2}}\xi^a\bar{K}_{a\mu\nu}\tau^{\mu\nu}
+ \frac{1}{2!}\frac{\omega^2}{\sigma}g_{ab}\xi^a\xi^b + {\cal
O}(\sigma^\frac{3}{2}),\label{1-37}
\end{eqnarray}
 where
\begin{eqnarray}\label{1-38}
\Box^{(0)} = -g^{ab}\partial_{a}\partial_{b},
\end{eqnarray}
and
\begin{eqnarray}\label{1-39}
\Box^{(1)} = -\frac{1}{|\bar{g}|^{1/2}}
\partial_{\mu}\bar{g}^{\mu\nu}|\bar{g}|^{1/2}\partial_{\nu}
+ \frac{1}{4}\bar{g}^{\mu\nu}\bar{g}^{\alpha\beta}
\left(
g_{ab}\bar{K}^{a}_{\,\,\,\,\mu\nu}\bar{K}^{b}_{\,\,\,\,\alpha\beta}
-
2g_{ab}\bar{K}^{a}_{\,\,\,\,\mu\alpha}\bar{K}^{b}_{\,\,\,\,\nu\beta}
\right).
\end{eqnarray}
\section{Wave functions and mass quantization}

Now using the equations (\ref{1-28})-(\ref{1-39}) and the ansatz
that $h_{\mu\nu}$ can be decompose as
$h_{\mu\nu}(x^{\mu},\xi)=\Sigma_{\beta}h^{(\beta)}_{\mu\nu}(x^{\mu})\Psi_{n}^{(\beta)}(\xi)$,
where the index $\beta$ runs over any degeneracy that exists in the
spectrum of the normal degrees of freedom, then  the wave equation
(\ref{1-29}) splits to the following two wave equations, normal and
tangent to the brane respectively
\begin{equation}\label{1-40}
\left(\Box^{(0)}+\frac{\alpha^{*}}{m-2}\omega^2
g_{ab}\xi^{a}\xi^{b}\right)\Psi_{n}^{(\beta)}(\xi)=\sigma
E_{n}\Psi_{n}^{(\beta)}(\xi),
\end{equation}
and
\begin{eqnarray}\label{1-41}
\left( -\frac{1}{|\bar{g}|^{1/2}}\partial_{\mu}
\bar{g}^{\mu\nu}|\bar{g}|^{1/2}\partial_{\nu} + {\cal D} -E_{n}- \frac{2(m-1)}{(m-7)(m-2)}\Lambda \right) h^{(\beta)}_{\mu\nu}(x^{\mu})
=0,
\end{eqnarray}
where
\begin{eqnarray}\label{1-42}
{\cal D} = \frac{1}{4}\left( g^{ab}\bar{K}_{a}\bar{K}_{b} -
2\bar{K}_{\alpha\beta a}\bar{K}^{\alpha \beta a}\right),
\end{eqnarray}
and
\begin{equation}\label{1-43}
\sigma
E_{n}=\left(\frac{2\alpha^{*}}{m-2}\right)^{\frac{1}{2}}\omega\sum_{n_=n_{1}}^{n_{m-4}}\epsilon_{n}\left(n+\frac{1}{2}\right),
\end{equation}
is the eigenfunction corresponding to the normal degrees of freedom,
describing $m-4$ uncoupled  harmonic oscillators. We know from
quantum mechanics that for harmonic oscillator only probability of
even modes around the minimum of potential is maximum, thus in this
model only even modes of wave functions are localized on the brane
($\xi^{a}=0$), in contrast to the usual brane models that only zero
mode of wave equation is localized on the brane \cite{17}.
It can be seen from equation (\ref{1-40}), without confining potential ${\cal V}$ the equation (\ref{1-40}) reduce to plane wave equation thus we can not localize  gravity on the brane. Hence not only we need confining potential
to localized test particles on the brane, but also it have the central role
in confining of gravitons and the amount of graviton mass.

According to the \cite{7} and \cite{dice}, we  assume $\alpha^*\omega^2 =
\frac{2}{\beta^2(m-2)}\Lambda^2$. Equations (\ref{1-41}) and
(\ref{1-43}) show that the mass of graviton is quantized
\begin{equation}\label{1-44}
M^{2}=\frac{2(m-1)}{(m-7)(m-2)}\Lambda\left(1+\frac{1}{\sigma}\,\,\sum_{n=n_{1}}^{n_{m-4}}\epsilon_{n}(n+\frac{1}{2})\right).
\end{equation}
In equation (\ref{1-44}) as $\sigma \rightarrow 0$, the mass becomes
very large. On the other hand, if we change the coordinates
according to $x^\mu \rightarrow \sqrt{\sigma}x^\mu$, then equation
(44) reduces to the following form
\begin{equation}
-\frac{1}{|\bar{g}|^{1/2}} \partial_{\mu}
\bar{g}^{\mu\nu}|\bar{g}|^{1/2}\partial_{\nu}
h^{(\beta)}_{\mu\nu}(x^{\mu})   - \sigma E_n
h^{(\beta)}_{\mu\nu}=0.\label{1-45}
\end{equation}
The above equation shows that the redefined mass is given by
\begin{equation}
\tilde{M}^2 = \sigma E_n = \frac{2(m-1)}{(m-7)(m-2)}\Lambda \sum
\epsilon_n (n+ \frac{1}{2}).\label{1-46}
\end{equation}
Since our change of coordinates amounted to $x^\mu \rightarrow
\sqrt{\sigma}x^\mu$, we relate this mass to the microworld. It is
the mass of a quantum perturbation in a spacetime with very small
curvature measured by the astrophysical value of $\Lambda$ as
opposed to the mass sometimes inferred from the zero point or vacuum
fields of particle interactions. As we noted above, the mass denoted
by $M$ as a consequence of the large-scale gravitational effects and
becomes very large according to $M = \tilde{M}/\sigma$. One may
interpret this as having a parameter relating the large scale to
that of the small. To have a feeling of $\sigma$, one may use the
fundamental constants to construct it. We assume that our brane has
a finite small thickness or width $l$. It also suggests to link the
squeezing parameter $\sigma$, determined by the constraining
potential, to the width of brane, that is
\begin{equation}
\sigma= al^b,\label{1-47}
\end{equation}
where $a$ is a dimensional constant so as to making $\sigma$
dimensionless so that one may use fundamental constants to determine
the constants $a$ and $b$. If we set $a = L^{-b}$ were $L =
\sqrt{\Lambda/3}$ is the intrinsic length scale of the universe and
if we set $b=2$ (this identification required from our rescaling
property of coordinates), then we have
\begin{equation}
\sigma = \left(\frac{l}{L}\right)^2.\label{1-48}
\end{equation}
According to the present experimental data in colliders, the value
of $l$ should be less or of order of $TeV^{-1}\sim 10^{-17} cm$. If
the brane width is $l$, it means that brane localized particles
probe this length scale across the brane and therefor the observer
cannot measure the distance on the brane to a better accuracy than
$l$. Hence we have
\begin{equation}
\sigma \sim \left(\frac{10^{-17}cm}{10^{28}cm}\right)^2
=10^{-90}.\label{1-49}
\end{equation}
 Now, inserting the appropriate units into equation (\ref{1-46}), we
obtain the fundamental mass $\tilde{M}$
\begin{equation}
\tilde{M}_0 = \frac{\hbar}{c}\Lambda^{\frac{1}{2}} \sim
10^{-65}gr.\label{1-50}
\end{equation}
Having made an estimate for $\sigma$, we can now obtain an order of
magnitude for $M_0$
\begin{equation}
M_0 \sim \left(\frac{L}{l}\right)^2 \tilde{M}_0 \sim 10^{25}
gr.\label{1-51}
\end{equation}
On the other hand, the observational data constrains the brane width
to be in order of planck length, see  \cite{dice} and references
therein. In this case we have \cite{7}
\begin{equation}
\sigma = \left(\frac{l_{Pl}}{L}\right)^2 = \frac{c^3}{\hbar G
|\Lambda|} \sim 10^{-120},\label{1-52}
\end{equation}
and consequently
\begin{equation}
M_0= \frac{\Lambda
c^4}{G}\frac{1}{|\Lambda|^{3/2}}\frac{1}{c^2}=\frac{c^2}{G|\Lambda|^{1/2}}
\sim 10^{56}gr,\label{1-53}
\end{equation}
 where $l_{Pl}$ is the Planck length. Let us ask about
the meaning that one should attribute to this $M$. The second
equality in (\ref{1-53}) contains three separate factors. The first
one represents the energy density generated by $\Lambda$; the second
term $|\Lambda|^{-3/2}$ is the total volume of the universe
restricted to its horizon and the last term is just to convert the
total energy into a mass. Thus we can interpret it as the total mass
of all existing gravitons in the observed universe. As was noted,
the squeezing parameter $\sigma$ was introduced as a consequence of
requiring a test particle to be confined to our brane. This
parameter, however, opens up an opportunity in the way of observing
our universe from two different angles, namely, either looking at
our universe in its entirety, that is, at its large-scale structure,
or do the opposite, i.e. observing it from small scales. The
parameter $\sigma$ provides us with the tool necessary to achieve
this. It suffices to define the change of variable introduced
earlier, that is $x^{'\mu}=\sqrt{\sigma}x^\mu$. This relation
connects two very different scales: small and large. Now the
question of the disparity between the values of the cosmological
constant in cosmology and particle physics reduces to its
measurement from two different scales. If we look at it from the
large-scale point of view, we measure its astrophysical value. On
the other hand, if one looks at its value from very small scales, it
turns out to be related to that of $\Lambda$ through the relation
$\tilde{\Lambda} = \sigma^{-2}\Lambda$ which is 240 orders of
magnitude larger than its astrophysical value. The unusually large
order of magnitude should not alarm the reader for if we had used
the Planck mass in equation (\ref{1-51}) instead of $M_0$, we would have
obtained the usual value for the order of magnitude, that is 120.
The above discussion leads us to the conclusion that the vast
difference between the values of the cosmological constant simply
stems from our measurements at two vastly different scales.
Note that in $4D$ generally covariant theory the mass is a scalar and is invariant
under any coordinate transformation. On the other hand, in this paper, we
introduced first a higerdimensional bulk space where the physical laws are
assumed to be covariant under general coordinate transformations and local
redefinitions of reference frames.
\begin{eqnarray}\label{1-54}
{\cal Z}^A\rightarrow {\cal Z}'^{A}({\cal Z}), \hspace{.5cm} E^A_I\rightarrow
\Lambda^B_I({\cal Z})E^A_B,
\end{eqnarray}
so that ${\cal G}_{AB}E^A_IE^B_J=\eta_{IJ}$. Also our assumption in the reduction
of dimensions of the bulk space to the brane is that the brane is parameterized
by coordinates $x^\mu$ and reference frames are made up $4$ reference vectors
$e^\mu_\alpha, \hspace{.3cm}\alpha=0,...,3$. $4D$ physical laws are covariant
under general coordinate transformations $x^\mu\rightarrow x'^\mu(x)$, see
for more details \cite{maraner}. The existence of a specified brane that we
are living in, dictates the broken of general covariance in the bulk space
\cite{maraner}. Consequently the re-scaling of graviton mass is a direct
result of the breakdown of general covariance in the bulk space. For a test particles
that are exactly confined to the brane, the influence of $4D$ general covariance
takes the mass of them invariant, but the graviton can be fluctuate in the
bulk space and consequently detects  the breakdown of $mD$ general covariance.

To make progress in our understanding of braneworlds, linear gravitational waves can also be most helpful.
Linear gravitational waves in the bulk space are supposed to propagate
both in the bulk and in the brane-world. One such propagation was considered
in \cite{Durrer}, with the surprising result that the the bulk wave interfered with the binary pulsar PSR1913+16 \cite{taylor},
producing a strong contrast with the Hulse-Taylor estimates on orbit decay.
In higher-dimensional theories where there are more than one extra dimensions the above result is not useable where in these models there are several gravi-scalars which may all contribute a similar amount to the emission of gravity waves \cite{Durrer}. 
The major exception  is that in warped models with non-compact
extra dimension such as Randall-Sundrum II model and our model, where the scalar gravity mode is not normalizable, this discussion dose not apply.
Also in this paper  we have studied a non-compact model without junction
conditions and therefor the calculations are not extendable to 
 our model.

\section{Conclusions} In this paper we have studied a perturbed
brane world model in which the matter is confined to the brane
through the action of a confining potential, rendering the use of
any junction condition redundant. In this approach, the mass of
graviton is quantized and the cosmological constant problem can be
addressed as the measurement problem. Also the spectrum of graviton
is discrete with localized even modes. The quantization of mass in
fact is the result of fluctuation of graviton around the original
non-deformed brane which is in agreement with results obtained in
Induced Matter Theory of P. S. Wesson \cite{wesson} and confined
quantized test particles in the branes \cite{7}.

\section*{Acknowledgements}
The authors thank  Prof. Sepangi for reading the manuscript.

\end{document}